\documentclass[doublecol]{epl2}
\usepackage{amssymb}
\usepackage{graphicx}

\title{
Charge Dynamics 
Across the Disorder Driven 
Superconductor-Insulator Transition
}

\author{Sabyasachi Tarat \and Pinaki Majumdar}
\shortauthor{Sabyasachi Tarat and Pinaki Majumdar}

\institute{
\inst{} Harish-Chandra  Research Institute,
Chhatnag Road, Jhusi, Allahabad 211019, India
}


\abstract{
While insensitive to weak non magnetic disorder, an $s$-wave 
superconductor can be driven insulating by strong disorder. 
Using a scheme that captures the correct 
ground state, and fully retains thermal amplitude and phase 
fluctuations, we describe the disorder driven superconductor-insulator 
transition at finite temperature. Our results on the resistivity
suggest that beyond moderate disorder the low temperature superconducting
state can arise out of an `insulating' normal state.
We also find that the low frequency weight in the density of states and 
optical conductivity are non monotonic in disorder, with a maximum
near critical disorder, and their high temperature values correlate
with the superconducting fraction in the disordered ground state.
}

\begin{document}

\maketitle

The disorder driven superconductor-insulator  transition (SIT) 
is a longstanding problem in condensed matter physics.
Although 
early theoretical work by Anderson \cite{anderson} suggested that 
superconductivity (SC) should be insensitive to 
non-magnetic disorder, a large number of experiments 
over the last couple of decades \cite{sit-revs} 
have revealed
that superconductivity is actually 
suppressed and finally destroyed by increasing disorder.
Simultaneously, the normal state resistivity changes
from metallic to insulating, and a pseudogap (PG) appears
in the single particle density of states.

The availability of high resolution scanning tunneling
spectroscopy (STS) tools has led to a significant advance
in the field. Recent observations
include:
i)~The increasing fragmentation of the SC state with disorder
\cite{sacepe-loc,sacepe-natc,sacepe-prl,pratap-ph-dg,chocka,pratap-prl,pratap-spat,noat}.
ii)~Survival of an apparent 
(pseudo)gap in the disorder driven normal state 
\cite{sacepe-loc,sacepe-natc,sacepe-prl,pratap-ph-dg}.
iii)~A change in the temperature dependence of the 
normal state
resistivity from metallic to insulating, without necessarily 
any universal temperature independent value at critical disorder 
\cite{baturina-res,kapit-res}.
Additionally, iv)~there are
observations of non monotonic magnetoresistance 
\cite{baturina-prl,baturina-res,adams-mr}, 
and finite frequency superfluid stiffness at large disorder
\cite{pratap-sf,crane-sf}, well past the SIT.
 
The fully self-consistent
Hartee-Fock-Bogoliubov-de Gennes (HFBdG) approach
\cite{ghosal} had already revealed that the
strong disorder SC ground state is fragmented in an essential
way, and predicted the survival of a single particle 
gap across the SIT.
Thermal effects have been probed using quantum Monte Carlo (QMC)
\cite{scal-dos,scal-res,bouad},  providing an
estimate of $T_c$ and the global 
density of states.  Surprisingly,  
apart from an early estimate \cite{scal-res},
there seem to be no results on the charge dynamics,
{\it i.e}, the resistivity and optical features,
across what is essentially a transport
transition.
The limitation stems from the inability to handle
thermal fluctuations on a large spatial scale, and
access real frequency information.

In this paper we solve this problem using
an approach that captures the HFBdG ground state, 
fully retains the thermal amplitude and phase fluctuations,
and locates the correct disorder scale, $V_c$, for
the zero temperature~SIT. 

Our main results, at intermediate coupling,
are the following:
i)~For $V \ll V_c$ the single particle gap closes
at $T_c$, but beyond $V \sim 0.25V_c$ there emerges a
pseudogap window above $T_c$, and when $V > 0.75V_c$ a
{\it hard gap} persists for $T > T_c$.
ii)~The normal state resistivity $\rho(T)$ is
`insulating', with $d\rho/dT <0$, already at $V \sim 0.5V_c$
so, for $0.5V_c < V < V_c$ one actually observes an
{\it insulator to superconductor} transition on cooling
the system.
iii)~Increasing temperature leads to a growth in the
low frequency single particle and optical weight
across $T_c$, over a window $\delta T \ll T_c$ at weak disorder
and $\delta T \gtrsim T_c$ at strong disorder. The $T \gg T_c$
weight  correlates closely with the `superconducting
fraction' in the ground state.
iv)~Increasing disorder leads to {\it non monotonic} behaviour
of the low frequency single particle and optical
weight, with a peak around $V_c(T)$.

While there is already experimental evidence for i) and ii),
the other
predictions can be tested experimentally and correlated with
spatial data where~available.

{\bf Model and method.} - We study 
the attractive Hubbard model in two dimensions (2D), with 
disorder incorporated in the form of a random 
potential $V_{i}$ at each site. 
\begin{equation}
H = H_{kin}
+ \sum_{i \sigma} (V_i - \mu) n_{i \sigma}
- \vert U \vert \sum_{i} n_{i \uparrow} n_{i \downarrow}
\end{equation}
where $H_{kin} =
- t\sum_{\langle ij \rangle \sigma}
c_{i \sigma}^{\dagger} c_{j \sigma}$.
$t=1$ denotes the nearest neighbour tunneling amplitude,
$U$ is the strength of onsite attraction, and 
$\mu$ the chemical potential, set so that the
electron density is $n \approx 0.9$.
The disorder $V_i$ is chosen from a box 
normalised distribution between $\pm V/2$.
We focus on the disorder dependence at 
$U=2t$ for specific results and 
discuss the $U$ dependence at the end.

We rewrite the interaction in terms of auxiliary `pairing'
and `density' fields via a Hubbard-Stratonovich (HS)
decomposition \cite{altland,erez},
and choose coefficients to recover HFBdG theory
at $T=0$.  We work in 
the static limit of the auxiliary fields, 
\begin{equation}
H_{eff}  = H_{kin}  + \sum_{i \sigma} (V_i - \mu) n_{i \sigma}
+ H_{coup} + H_{cl}
\end{equation}
where $H_{coup} = \sum_i 
(\Delta_i c^{\dagger}_{i\uparrow} c^{\dagger}_{i \downarrow}
+ h.c) + \sum_i \phi_i n_i$ and $H_{cl} =
(1/U)\sum_i (\vert \Delta_i \vert^2 + \phi_i^2)$.
The presence of the 
field $\phi$ is important, especially in disordered systems, 
since it captures the enhancement of disorder due
to interaction effects. 

Unlike mean field theory, we {\it do not minimise} the free
energy with respect to $\Delta_i$ and $\phi_i$ at 
finite temperature, $T$, but sample
all configurations according to their Boltzmann weight
$ P\{\Delta, \phi\} \propto Tr_{c,c^{\dagger}}
e^{-\beta H_{eff}} $. The difficulty in this approach
is of course in evaluating the weight for an
arbitrary configuration. This is handled using 
the Metropolis
algorithm combined with a diagonalisation
of the effective Bogoliubov-de Gennes (BdG) problem
\cite{dubi,dag}.
We use an ${\cal O}(N)$ cluster based scheme 
\cite{tca} for the Monte Carlo updates and can
readily access system sizes $ \sim 24 \times 24$. 
Our method captures the HFBdG ground state as $T \rightarrow 0$,
since the $\Delta_i$ and $\phi_i$ minimise the free energy in
that case, but captures the crucial amplitude and phase
fluctuations at finite temperature.
We have focused on $U =2t$, which is the weakest interaction
we can controllably access 
with our system size. Even at $U=2t$ the
gap to $T_c$ ratio in the clean limit
is already outside the BCS window. 
Our results are averaged over atleast 10 copies of disorder.

\begin{figure}[t]
\centerline{
\includegraphics[width=6.0cm,height=4.3cm,angle=0]{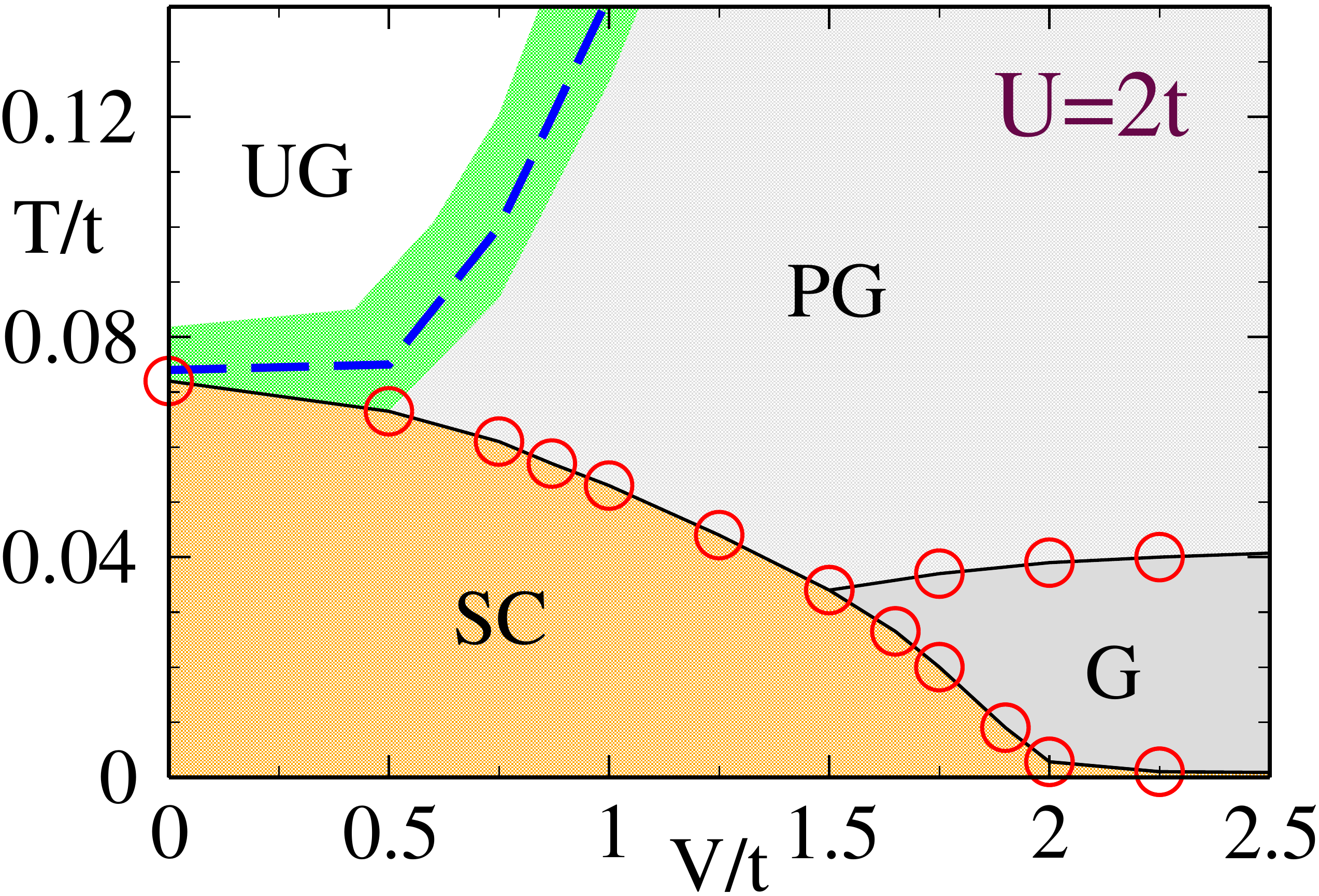}
}
\caption{(Colour online) Phase diagram for $U=2t$ showing the
superconducting (SC), and the following non superconducting
phases: gapped (G), ungapped (UG) and
pseudogapped (PG).
We take the critical disorder $V_c \sim 2t$. The tail shows the
exponentially small superconducting $T_{c}$ surviving beyond $V_{c}$.
A normal state pseudogap shows up for $V \gtrsim 0.25 V_c$ and for
$V \gtrsim 0.75 V_c$ the $T \gtrsim T_c$ phase actually has a
hard gap.
The crossover between pseudo-gapped
and notionally ungapped phase is shown by the green area. The blue
dashed line shows the transition from an `insulating'
($d\rho/dT < 0$) to `metallic' regime, which lies within the
broad crossover.}
\end{figure}

{\bf  Phase diagram.} -
We begin with the $V-T$ phase diagram  
in fig. 1, detailing 
the different phases in terms of transport and spectral
character. In the absence of disorder SC order is lost
at $T_c = T_c^0 \sim 0.07t$, benchmarked with QMC results \cite{rand-tc}. 
Increasing disorder leads initially to a slow suppression of $T_c$
which accelerates for $V >  1.5t$.
Our method reduces to HFBdG at $T=0$ so
the ground state within our scheme is 
in principle always superconducting. 
However, the phase
stiffness of the HFBdG state 
reduces quickly with increasing
disorder and for $V \gtrsim 2t$ we do not observe a 
transition down to $\sim 0.05 T_c^0$. 
We indicate the disorder scale associated with this
resolution limit on $T_c$ as $V_c$. 
Figure 1 indicates the exponentially small $T_c$
that survives in principle beyond our $V_c$. This would
be pushed to zero by quantum phase fluctuations in the
$\Delta_i$. 
$T_c^0$ and $V_c$ will set the natural scales of temperature
and disorder for us.
Our observations are the following.

i)~At weak disorder the fractional 
suppression of $T_c$ is small, {\it e.g},
at $V = 0.25 V_c$, $T_c$ falls less than $10\%$ from
$T_c^0$. 
This is qualitatively consistent with the 
`insensitivity' to disorder predicted by
Anderson \cite{anderson}, 
but that approach fails to be useful
beyond weak disorder. 
We are not aware of analytic results on the suppression of $T_c$
in this coupling regime, although there are numerical results on the
reduction of the phase stiffness \cite{ghosal}.
Overall, the weak disorder regime, $ 0 < V < 0.25V_c$, 
corresponds to almost homogeneous $\vert \Delta_i \vert$
in the ground state, metallic resistivity above $T_c$, and 
no significant anomaly in the normal state density of 
states (DOS).

ii)~The intermediate disorder window, $0.25V_c < V < 0.75V_c$,
resists easy characterisation. $\vert \Delta_i \vert$ in the
ground state shows increasing fragmentation \cite{ghosal,unpub}.
The resistivity is `insulating' near $T_c$ and
crosses over to metallic behaviour at high $T$,
while the $T > T_c$ density of states shows a pseudogap.  

iii)~At strong disorder, $V \gtrsim 0.75 V_c$, the
pairing amplitude $\vert \Delta_i \vert$ 
 is very inhomogeneous in the ground
state, a hard gap survives in the DOS even above $T_c$,
and the resistivity shows activated behaviour. This is
a regime where the low $T$ superconducting state
emerges from a high $T$ {\it insulating} phase, suggesting
that the $T_c$ is no longer controlled by the low energy DOS.

The different phase boundaries would depend on $U/t$.  
Apart from the change in $T_c^0$ and $V_c(T=0)$,
increasing $U/t$ would lead to an increase in the `gapped'
region, while decreasing $U/t$ would decrease the gapped
window in favour of the PG (and the UG to PG
crossover could be pushed to larger $V/V_c$).

%
\begin{figure}[t]
\centerline{
\includegraphics[width=4.2cm,height=3.1cm,angle=0]{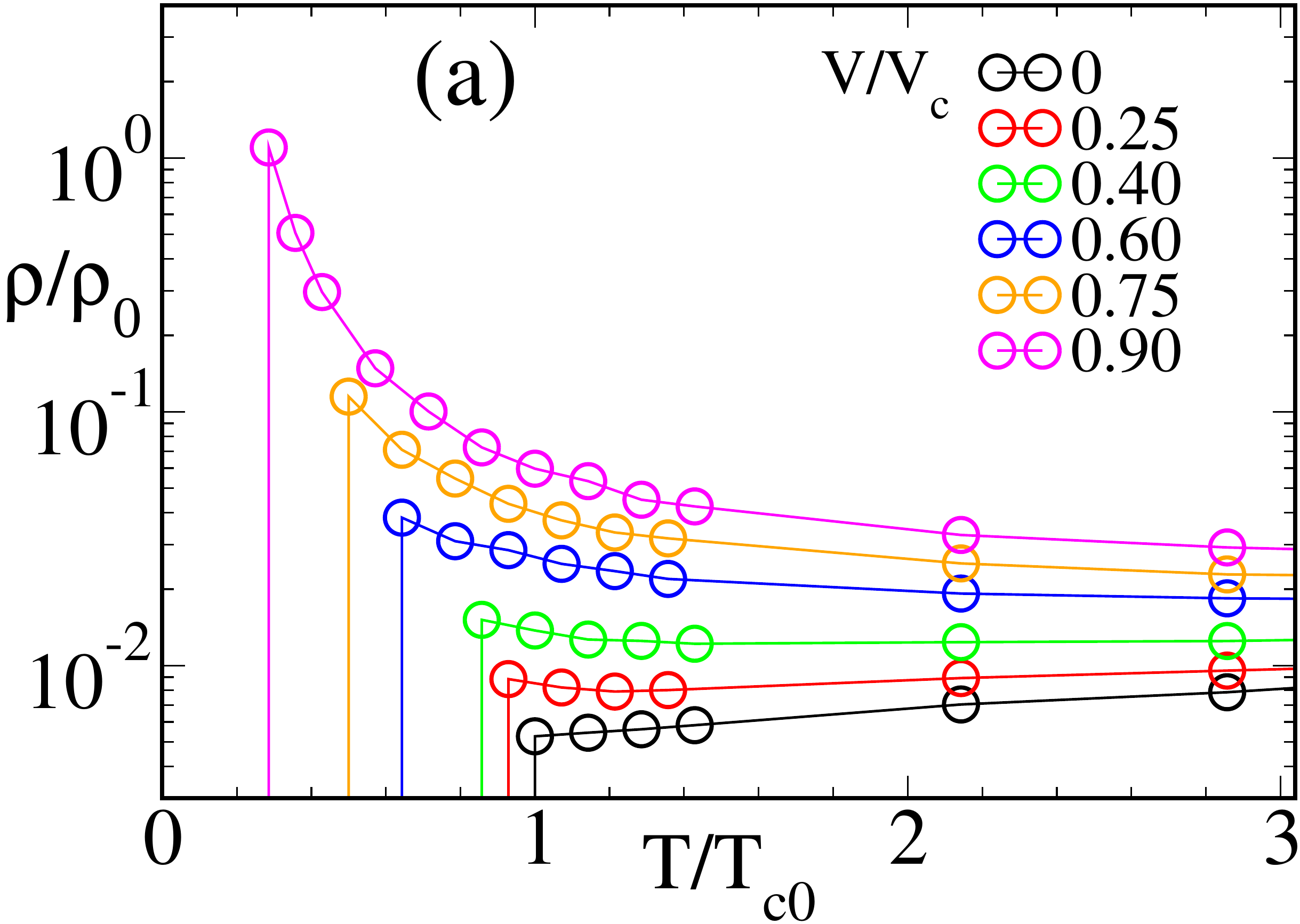}
\includegraphics[width=4.2cm,height=3.1cm,angle=0]{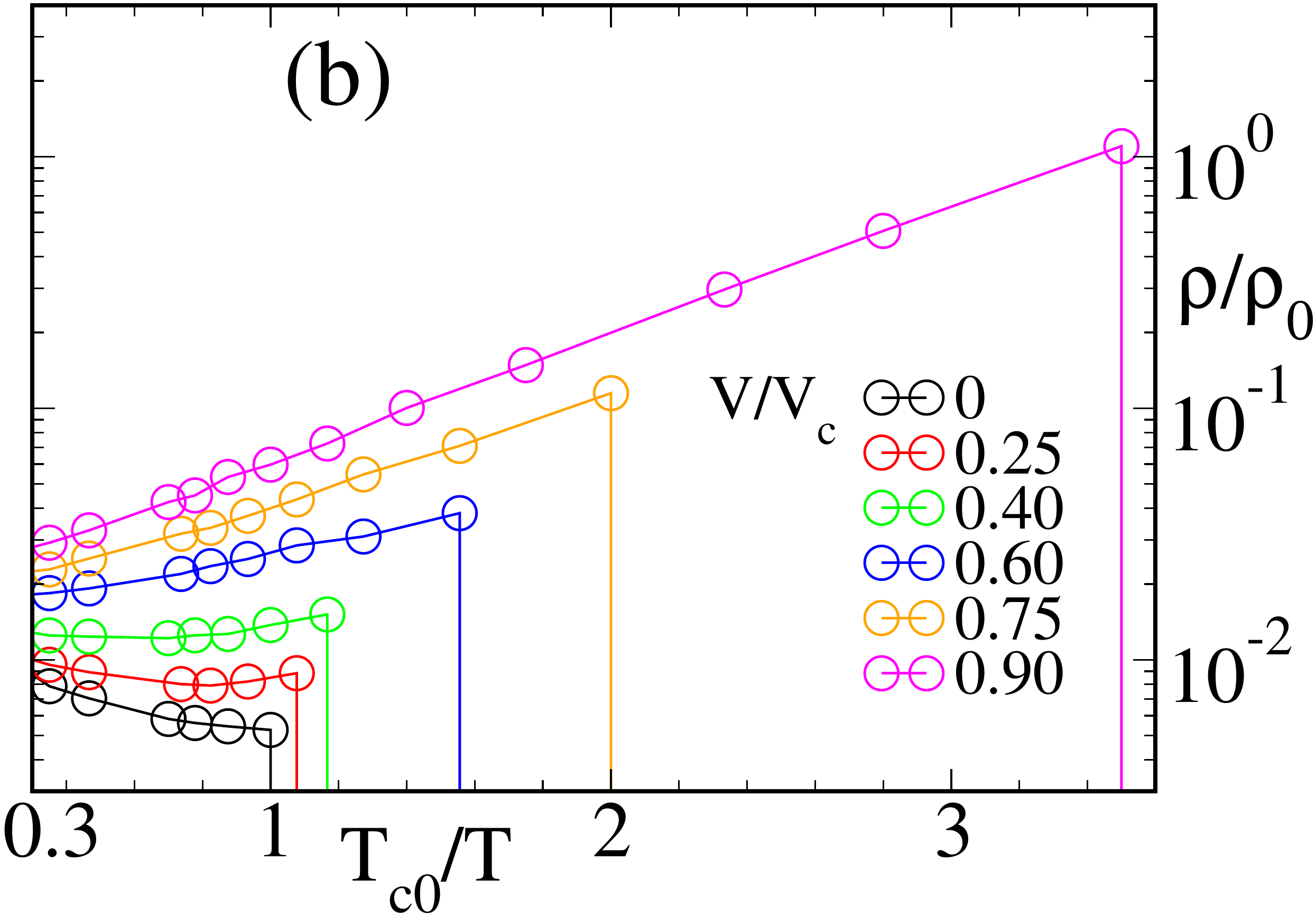}
}
\caption{ (Colour online)
(a) The resistivity, $\rho(T)$, measured in units of 
$\rho_{0} = {\hbar}/(\pi e^2)$,
evolving from metallic to insulating behaviour in the
normal state with growing disorder. For $V \lesssim 0.25V_{c}$,
it is metallic, between $0.25V_{c} \lesssim V \lesssim 0.75V_{c}$,
it is mixed, showing a thermal transition from `insulating' at low $T$
to weakly `metallic' at larger $T$. Beyond $V=0.75V_{c}$, the low $T$
behaviour is exponential $\rho(T) \propto e^{\Delta_g/T}$, with 
$\Delta_g$ increasing with $V$. This is highlighted in (b), where we see
that such a fit ceases to be valid below $V \sim 0.75 V_{c}$.
}
\end{figure}

{\bf Resistivity.} -
The resistivity
$\rho(T)$ is computed using the Kubo formula, via the
low frequency limit of the optical conductivity $\sigma(\omega)$.
Formally
$ \sigma(\omega) = -\omega^{-1} Im(\Lambda_{xx}(q=0,\omega)) $
where the current-current correlation function is defined by
$$
\Lambda_{xx}(q=0, \omega) =
\frac{1}{ \cal{Z}} \sum_{n,m} | \langle n | j_{xx} | m \rangle |^{2}
\frac{e^{- \beta E_{n}}- e^{- \beta E_{m}}}{\omega + E_{n}- E_{m}+ i \delta}
$$
The $\vert n \rangle$, $\vert m \rangle$
are many particle eigenstates of the system,
$j_{xx}$ is the current operator.
For the regular part,  $\sigma_{reg}(\omega)$,
{\it i.e}, excluding the superfluid response, it simplifies
within our static auxiliary field theory to:
\begin{eqnarray}
\sigma_{reg}(\omega) & =&
\sum_{a,b} F_1(a,b)
\frac{(n(\epsilon_{a}) + n(\epsilon_{b}) - 1 )}
{\epsilon_{a} + \epsilon_{b}}
\delta(\omega - \epsilon_{a}- \epsilon_{b}) \cr
&&~+ \sum_{a,b} F_2(a,b)
\frac{(n(\epsilon_a ) - n(\epsilon_{b}) )}{\epsilon_{a} - \epsilon_{b}}
\delta(\omega - \epsilon_{b} + \epsilon_{a})
\nonumber
\end{eqnarray}
where, now, the $\epsilon_{\alpha}, \epsilon_{\beta}
> 0$, {\it etc}, are {\it single particle
eigenvalues} of the BdG equations, the $n(\epsilon_a)$, {\it etc.},
are Fermi functions, and the $F$'s are current matrix elements
computed from the BdG eigenfunctions.
The dc resistivity is defined for $T > T_c$
via  $\rho^{-1} =  {\omega_0}^{-1}
\int_0^{\omega_0}  \sigma_{reg}(\omega) 
d \omega$, where $\omega_0 \sim 0.1t$.

Figure 2.(a) shows $\rho(T)$ 
for different $V$.
We use $d\rho/dT > 0$ to indicate a metal and
$d\rho/dT < 0$ to indicate an insulator.
Upto $V \sim 0.25V_c$ 
the resistivity is metallic at all $T$,
 except very near $T_c$. For 
$0.25 V_c < V < 0.75V_c$, however, the behaviour is mixed, with `metallic'
character at high $T$ and an `insulating' window below.
As the companion plot, fig. 2(b), shows the weakly
insulating behaviour cannot be characterised by a $T$
independent gap in the DOS. 
Beyond $0.75 V_c$ the resistivity is insulating at all $T$,
we have checked it upto $T=0.3t$. The log plot in fig. 2(b)
shows that the resistivity can be modeled as $\rho(T)
\propto e^{\Delta_g/T}$, with 
a weakly disorder dependent coefficient $A$ 
and an activation scale $\Delta_g \sim 2.5(V - 0.75 V_c)$.

The large disorder regime admits a simple explanation
in terms of the current paths. 
The low energy excitations are localised in the
superconducting clusters that form at $T=0$.
Since the SC regions are `disconnected' at $T > T_c$
all current 
paths have to pass partly through the insulating matrix,
leading to an activation factor in the conductance.
At weaker disorder the SC clusters
have a tenuous connection and the detailed frequency
dependence of the DOS is important.

\begin{figure}[t]
\centerline{
\includegraphics[width=4.1cm,height=3.1cm,angle=0]{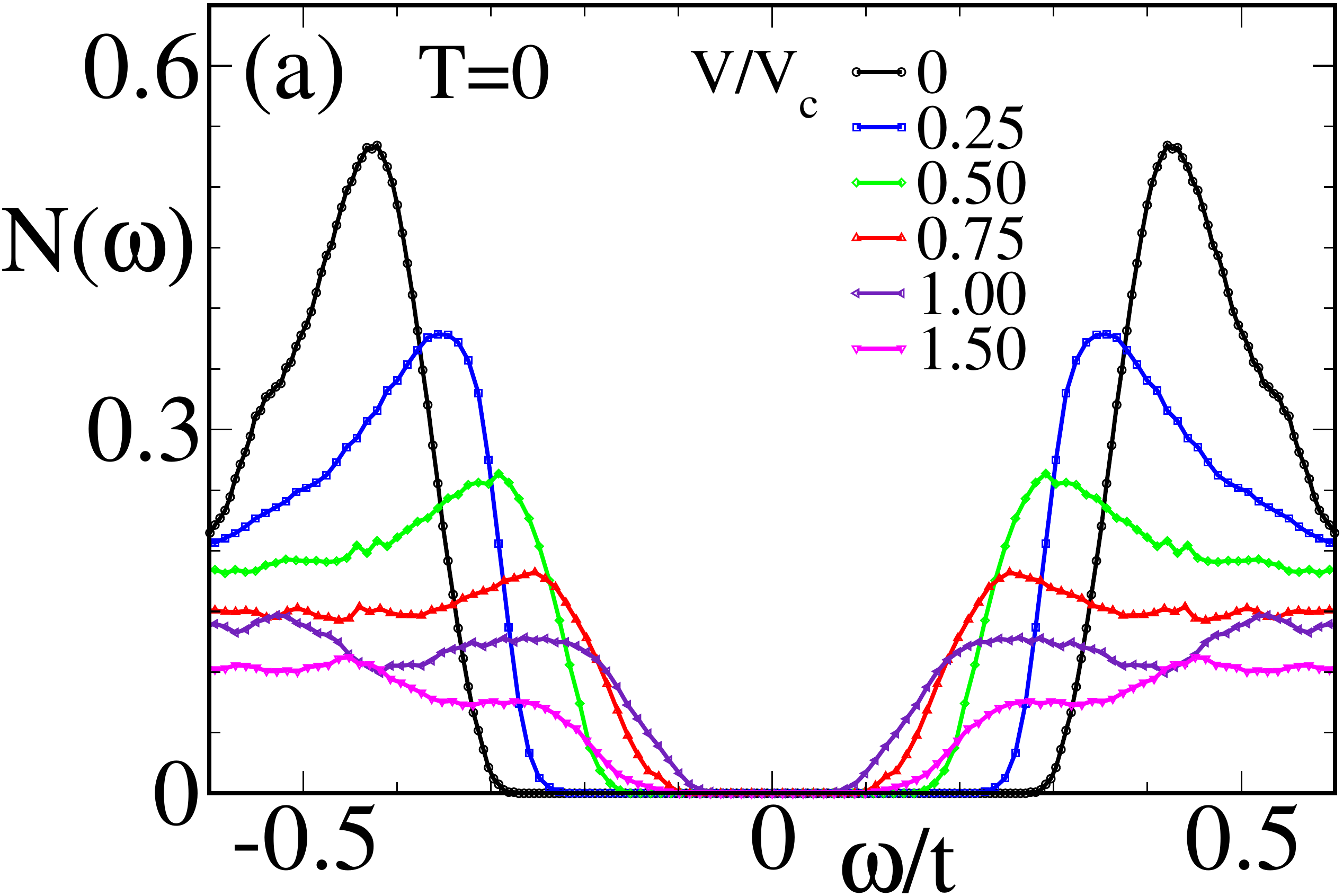}
\includegraphics[width=4.1cm,height=3.1cm,angle=0]{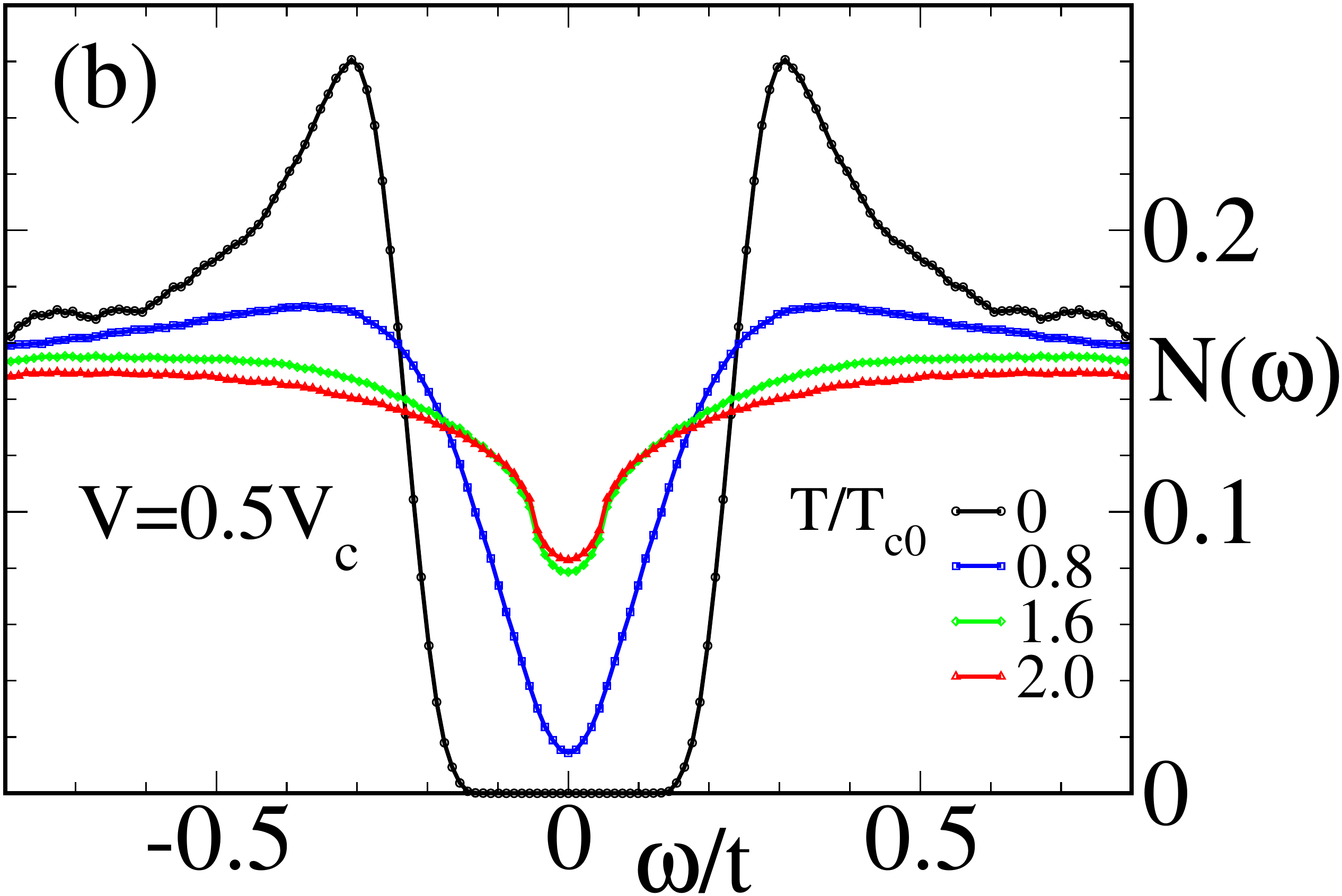}}
\vspace{.1cm}
\centerline{
~~\includegraphics[width=4.1cm,height=3.1cm,angle=0]{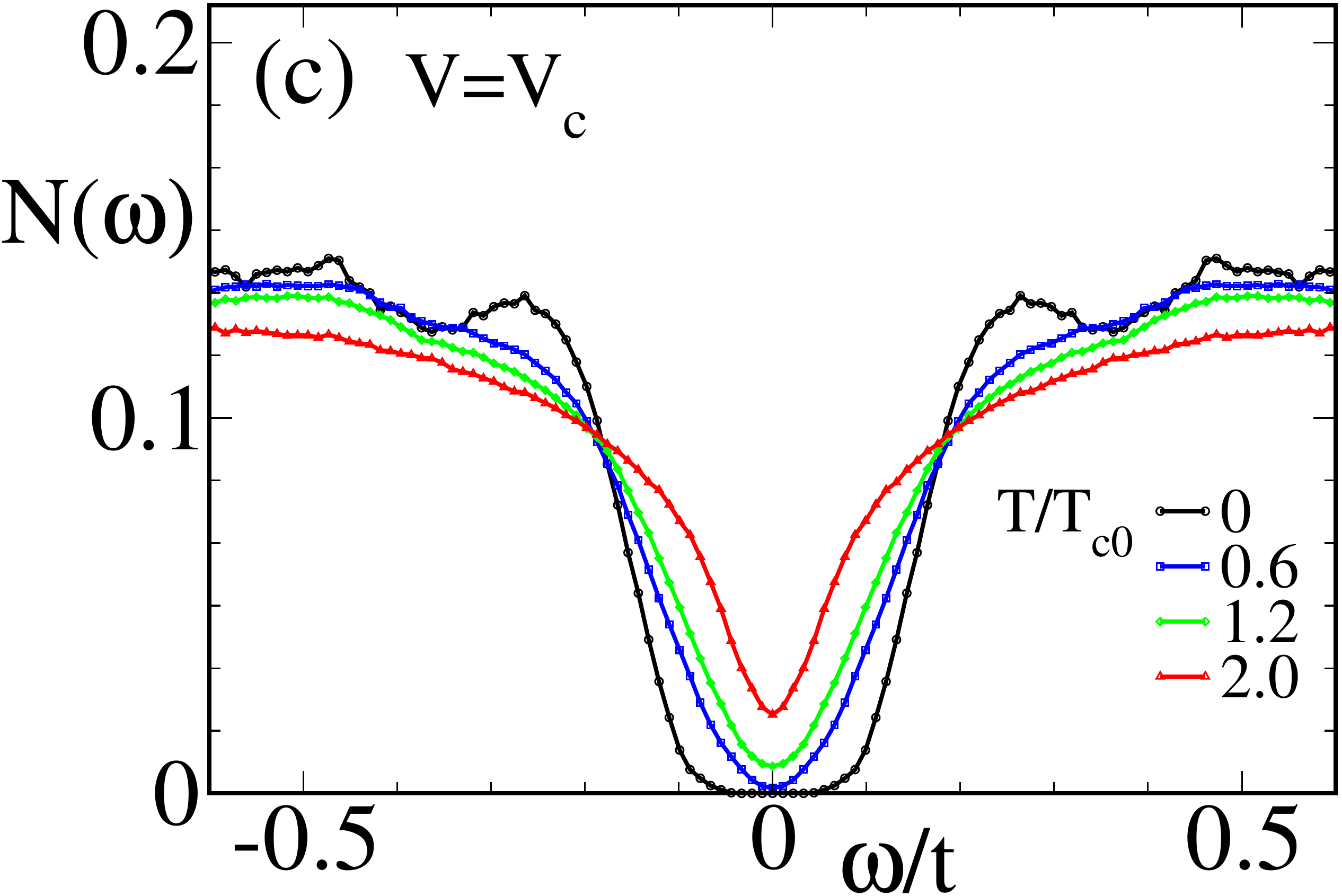}
\includegraphics[width=4.1cm,height=3.1cm,angle=0]{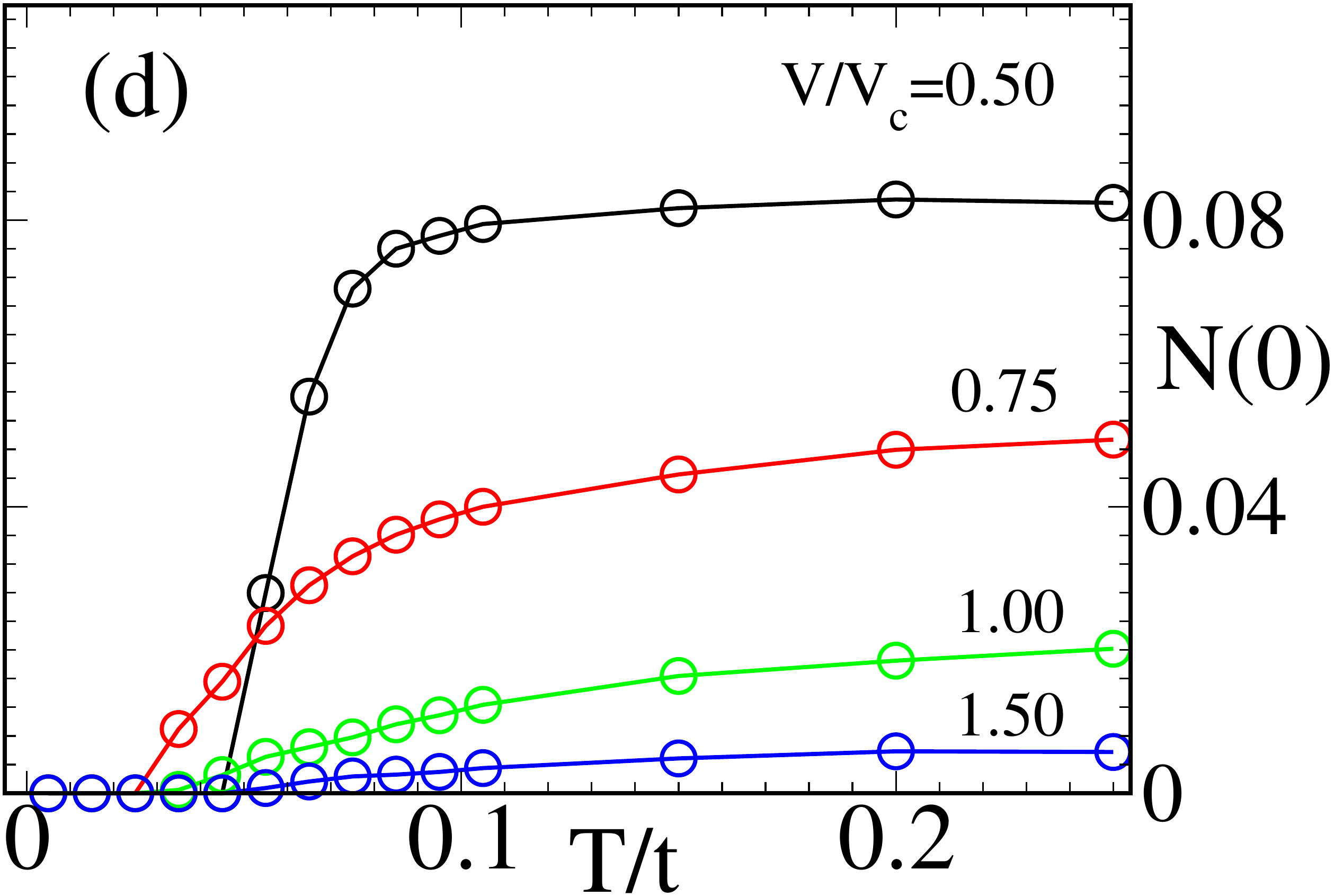}
}
\caption{(Colour online) Density of states at $U=2$.
(a)~The DOS at low temperature, showing the persistence of a
gap at all $V$, while the coherence peaks are difficult
to discern beyond $V \sim 0.75V_c$.
(b)~Temperature dependence of the
DOS for $V = 0.5V_c$, 
already showing a noticeable pseudogap
for $T > T_c$.
(c)~Same as (b) but for $V \sim V_c $, where the
system is insulating at all temperature.
(d)~Temperature dependence of $N(0)$, the
DOS at the Fermi level, for different disorder.
}
\end{figure}

Overall, we observe 
that for our chosen  $U=2t$, superconductivity
can arise out of a 
metallic or an insulating normal state. 
Our calculation does not hint at any `universal'
temperature independent resistance at $V=V_c$,
apparently consistent with
recent experimental analysis \cite{baturina-prl}.

{\bf Density of states.} -
Fig.3 shows the variation in
the single particle DOS with disorder and temperature.
If $\epsilon_n$
and $\{ u^{i}_{n},v^{i}_{n} \}$  are the positive
BdG eigenvalues and eigenvectors, respectively,
 in some equilibrium
configuration, the DOS is computed as 
$$
N(\omega) = \langle \sum_{i,n} |u^{i}_{n}|^{2}
\delta(\omega - \epsilon_n)
+ |v^{i}_{n}|^{2} \delta(\omega + \epsilon_n)\rangle 
$$ 
where the angular brackets indicate thermal average.

In fig. 3(a) we show the DOS plots at $T=0$ for 
varying disorder, from $V=0$ to $V=1.5V_c$.
The following features are noteworthy. 
(i)~The system is gapped for all $V$. 
The `gap' shows a non-monotonic character, decreasing upto 
$V \sim V_c $ and increasing from thereon. These results
match well with previous BdG \cite{ghosal}
and QMC \cite{bouad}  benchmarks.
(ii)~The coherence peaks 
decrease with increasing $V$, and beyond $V \sim 0.75V_c $ they are
hard to discern in the DOS. The `rise' of the DOS at the
gap edge is ideally
sharp in the clean limit, corresponding to the square root
BCS singularity, but for $V \gtrsim 0.75V_c $ the rise is
much gentler. 

Figure 3(b) 
shows the thermal evolution at $V=0.5V_c$, 
intermediate disorder, 
while 3(c) shows the same at 
$V=V_c$.
For $V=0.5V_c$ the coherence peak 
at the gap edge vanishes at $T \sim T_c \approx 
0.8T_c^0$, the low frequency DOS grows steadily 
with increasing temperature, but
a pseudogap feature surives upto $T \sim 2.5T_c$.
Figure 3(c) shows the result at $V=V_c$, where
the system retains a hard gap with increasing 
temperature
till $T \sim 0.5T_c^0$ and a deep pseudogap thereafter.

Figure 3(d) shows the temperature dependence of 
$N(0)$, the DOS at the Fermi level, for a few $V$. At
$V = 0.5 V_c$, $N(0)$ is essentially zero till 
$T \sim T_c$ and then rises quickly and saturates to 
a high $T$ asymptote.  With growing disorder the 
temperature interval $\delta T(V)$
 over which the rise occurs increases
and the `asymptotic' high temperature value
reduces. We find that this high temperature 
value at a given disorder  roughly corresponds
to the superconducting fraction 
in the ground state \cite{unpub} at that 
disorder. 

{\bf Optics.} -
Figure 4(a)-(c) shows aspects of 
the optical conductivity 
at two representative temperatures 
for various $V$, while 4.(d) shows the
integrated low frequency 
weight as increasing disorder drives the 
SIT.

In a disordered 
superconductor, the real part of the 
optical conductivity
consists of a delta function at $\omega = 0$
(signifying dissipationless transport) and a 
`regular' part $\sigma_{reg}(\omega)$. 
Within a mean field picture,
$\sigma_{reg}(\omega)$ at $T=0$ 
is suppressed for  $\omega \lesssim 2 \Delta$,
the gap scale. Beyond this
$\sigma_{reg}$  
rises to a peak and for $\omega \gg \Delta $
tends to the disordered metal limit, 
$ \Gamma/(\omega^{2}+\Gamma^2)$,
where $\Gamma$ is the scattering rate.
Figure 4(a) shows the low $T$ result at different $V$,
consistent with this general expectation. The low $V$
curves have a gap, rise to a relatively sharp
maximum and then fall off. Increasing $V$ increases 
$\Gamma$ making the fall off broader. 
In what follows we use just $\sigma(\omega)$ 
rather than $\sigma_{reg}(\omega)$ to denote
the regular part.

It is useful to compare these results with that
of Mattis-Bardeen (MB) theory \cite{mbtheory}, 
which is formulated
in the weak coupling limit. 
Qualitatively, within MB theory 
the thermal excitation of quasiparticles to
the gap edge, with a probability $\sim e^{-\Delta(T)/T}$,
leads to a `subgap' feature in $\sigma(\omega)$ at
finite $T$. Due to the large DOS at the gap edge
this contribution to $\sigma(\omega)$ 
is large at low $\omega$. Disorder broadens the 
coherence peaks and makes this subgap $\omega$
dependence flatter.

Figure 4(b) shows the low frequency results for       
$\omega \lesssim 2 \Delta_{0}$ at  $T=0.2T^0_c$. 
In this frequency range, $\sigma(\omega)$ decreases
monotonically with decreasing frequency, forming
a hard gap at a disorder dependent frequency $\omega_{g}(V)$.
$\omega_g$ is lowest between $0.75V_{c}$ and $V_{c}$, 
which seems to match with the critical
disorder $V_{c}(T)$ at this $T$, see Fig.1.
We do not find any  discernible subgap feature at 
this temperature. 
The thermal factor $e^{-\Delta/T} \sim 
e^{-{4 T_c^0}/{0.2T_c^0}}$ is too small.

At $T=0.7T_c^0$, fig. 4(c), the
cleaner samples do show an upturn reminiscent of
MB theory. However, even here for $V \gtrsim 0.5V_c$
the low frequency peak is absent due to the 
disorder and temperature induced broadening of the
coherence peak.

In Figure 4(d), we show the the low frequency 
optical weight $w_{opt}(V,T,\Omega) = \int_0^{\Omega}
d\omega \sigma(\omega,V,T)$ with $\Omega = 0.2t$.
We find that the maximum in $w_{opt}(V,T)$ at a given $T$ occurs at
a disorder $V_{max}(T)$ that tracks 
the critical disorder $V_{c}(T)$. The inset shows 
the low frequency single particle weight
$w_{dos}(V,T) = \int_0^{\Omega/2}
d\omega N(\omega,V,T)$ and its
strong correspondence with $w_{opt}(V,T)$.
At $T=0$ the behaviour of $w_{dos}$ is 
consistent with earlier observation \cite{ghosal}
that the gap is minimum close to $V_{c}$.
The resulting maximum in $w_{dos}$ persists, surprisingly, 
at finite $T$ as well and also shows up in the behaviour of
$w_{opt}$.


\begin{figure}[t]
\centerline{
\includegraphics[width=4.1cm,height=3.1cm,angle=0]{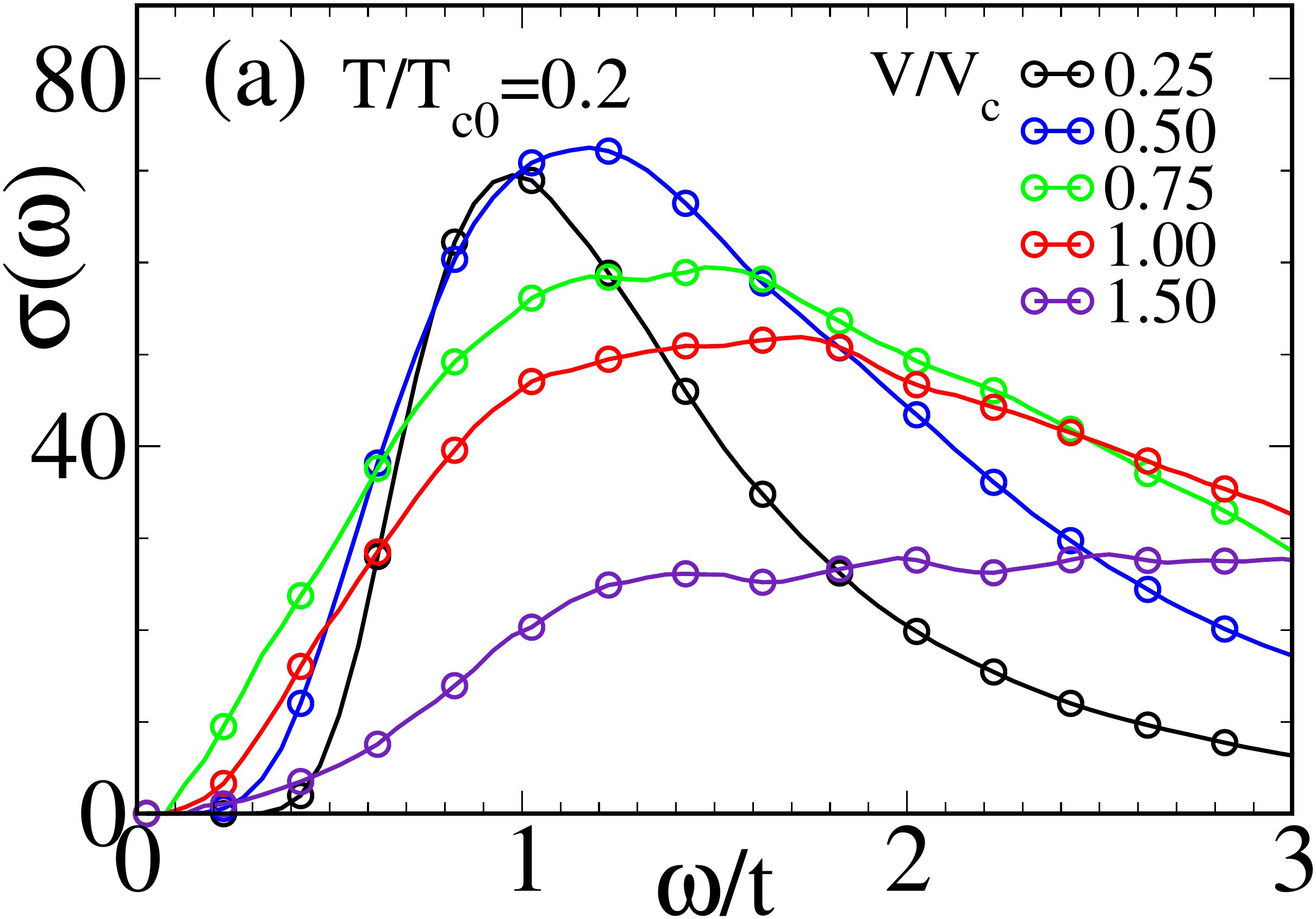}
\includegraphics[width=4.1cm,height=3.1cm,angle=0]{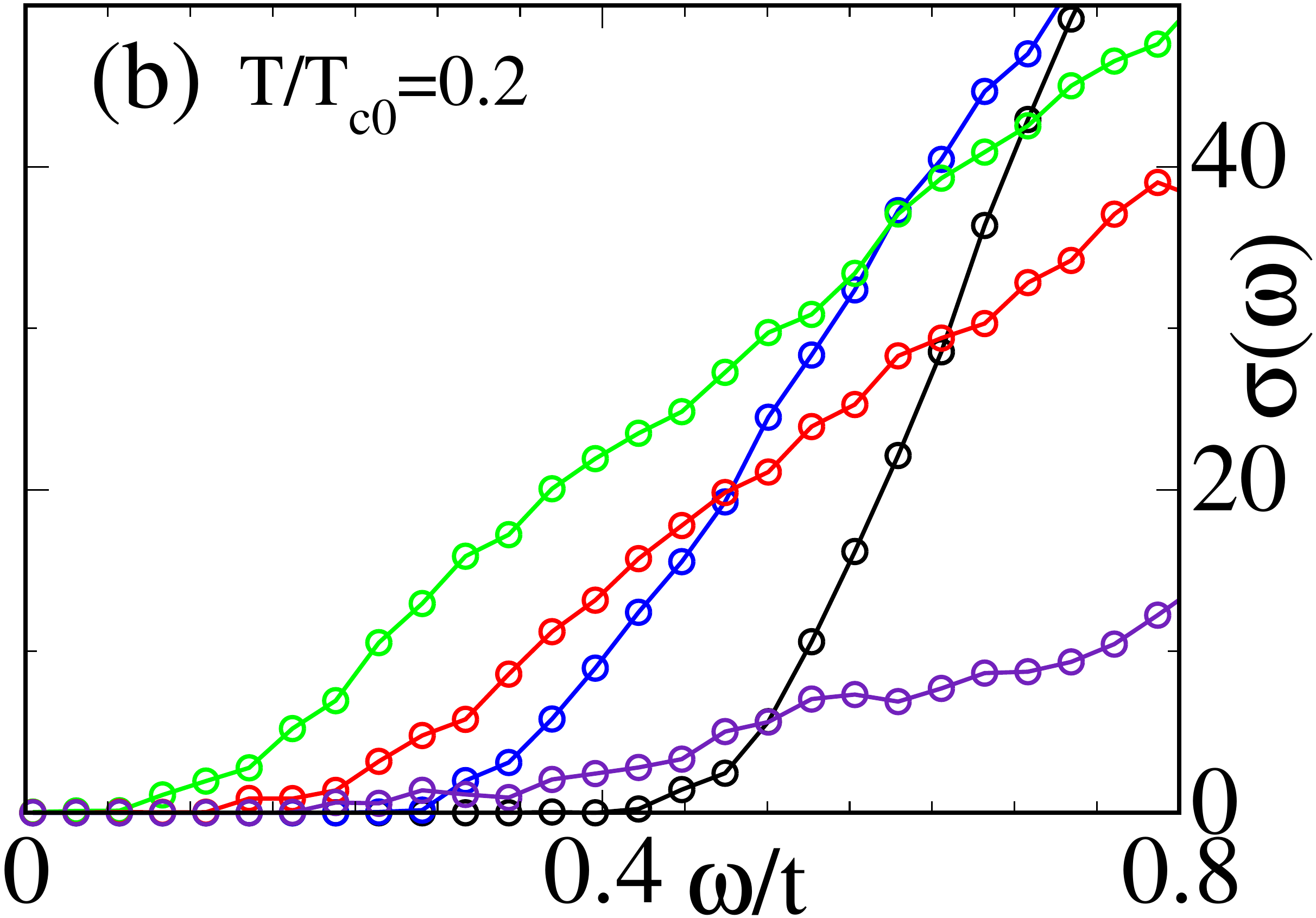}}
\vspace{.1cm}
\centerline{
\includegraphics[width=4.2cm,height=3.1cm,angle=0]{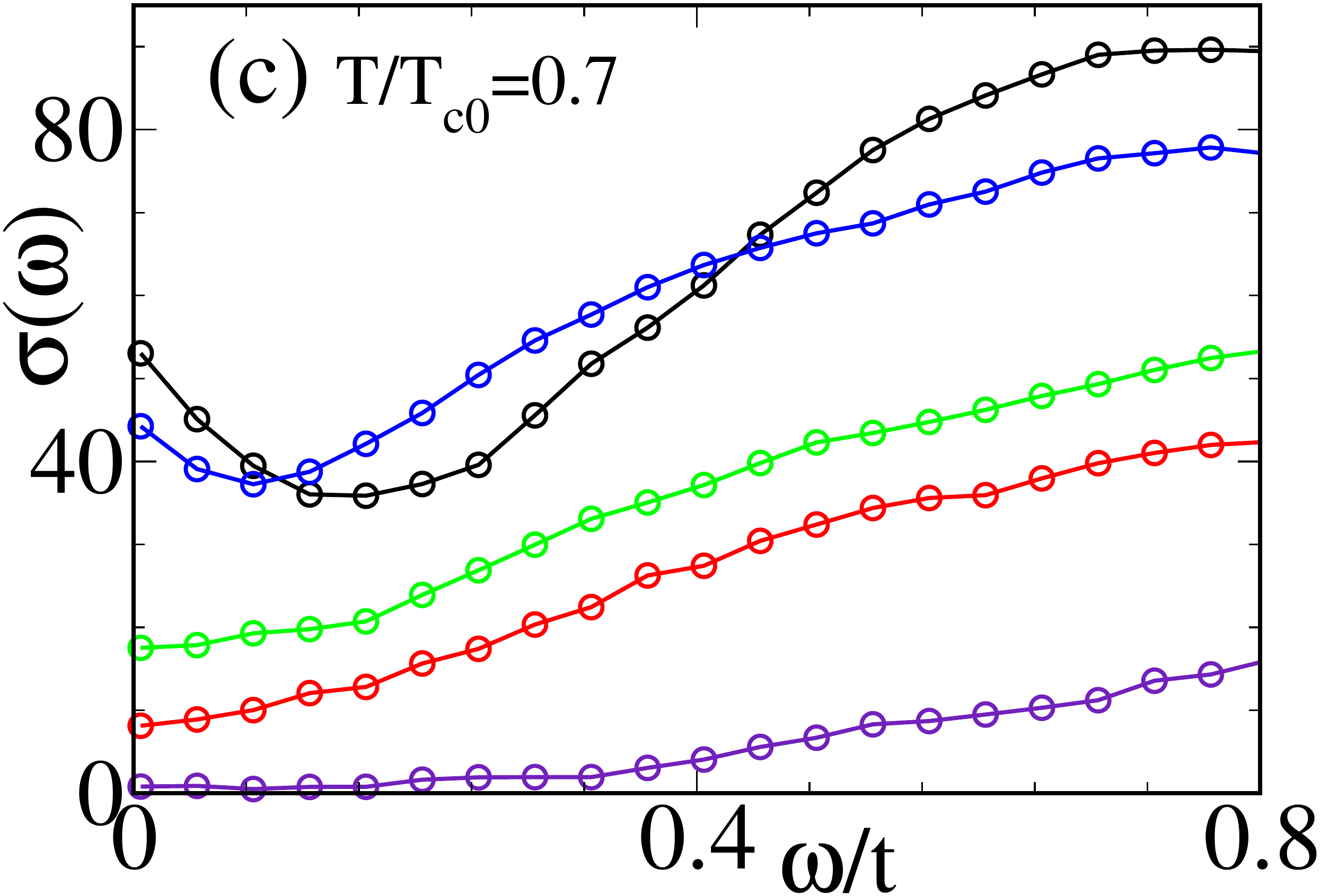}
\hspace{-.2cm}
\includegraphics[width=4.1cm,height=3.1cm,angle=0]{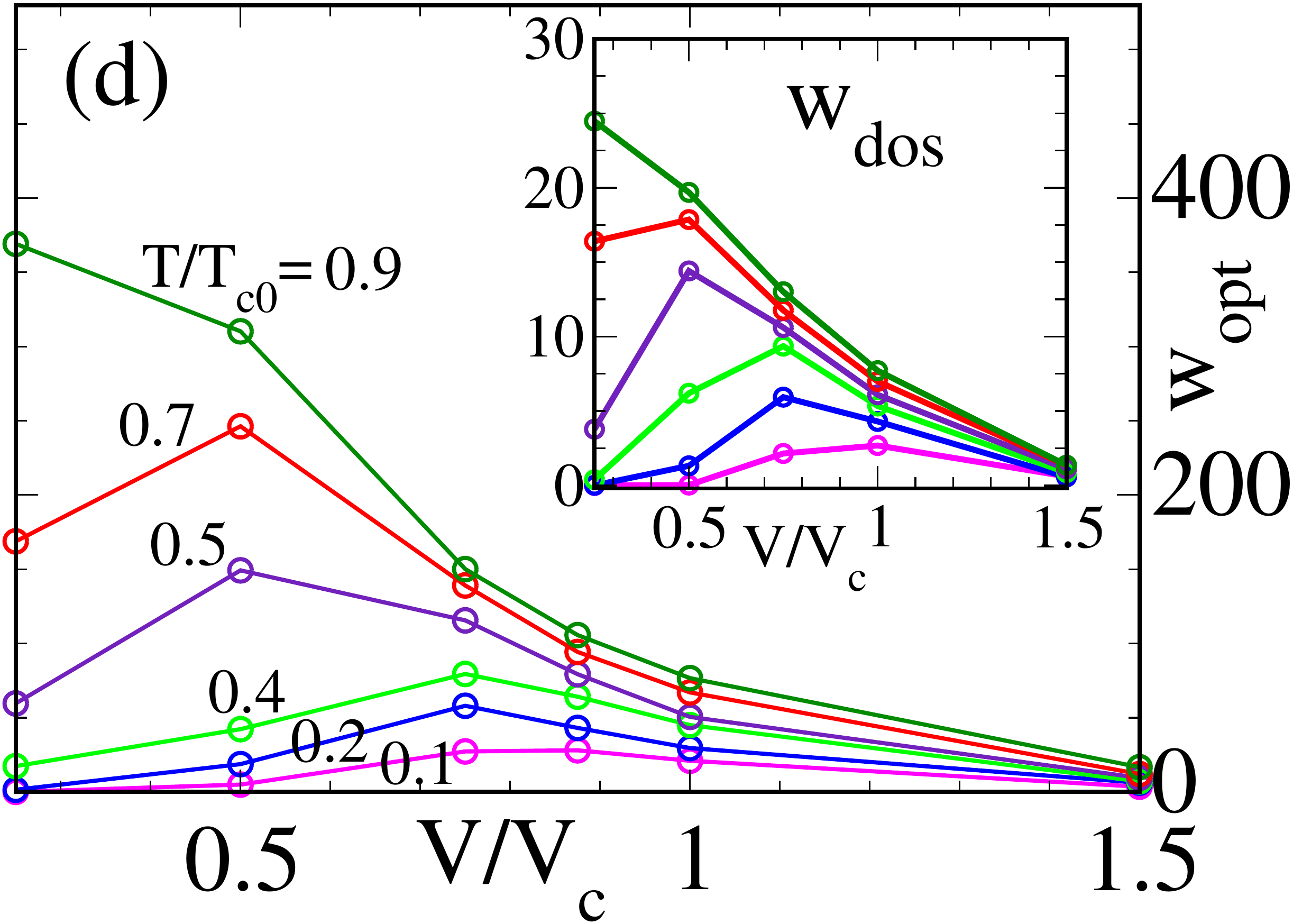}}
\caption{(Colour online)
Optical conductivity and low frequency optical spectral weight.
(a)~The behaviour of
$\sigma(\omega)$, measured in units of $\sigma_{0} = \pi e^2/{\hbar}$,
over a wide frequency range
for disorder varying across the SIT. The temperature is
$T=0.2T^0_c$.
(b)~The low frequency behaviour of $\sigma(\omega)$ for
varying $V$ (same legends as in panel (a)), at $T=0.2T_c^0$.
  (c)~Same as in (b), now at $T= 0.7T_c^0$. Notice the absence of any
gap, and the low frequency upturn,
in samples with $V=0.25V_c$ and $V=0.5V_c$ which are still
below their respective $T_c$.
(d)~Disorder dependence of the low frequency optical
spectral weight, $w(V,\Omega)$, see text, at different $T$. Inset shows
low frequency weight of the single particle spectrum.
}
\end{figure}

Since the weights are readily measurable, a qualitative
explanation may be useful.
Within our scheme the optical spectrum arises as
a convolution over the single particle Greens function,
so understanding $w_{dos}(V,T)$ can shed light on
$w_{opt}$ as well. Figure 3(d) provides a hint, where,
crudely, $N(0,V,T)$ rises from zero at $T \ll T_c$ to its
high $T$ asymptote, $N_{\infty}(V)$, say, across $T_c$.
$N_{\infty}(V)$ reduces monotonically with $V$.
If we ignore the `width' $\delta T$ of the low to
high temperature transition, then $N(0,V,T) \approx 0$ 
for $T <  T_c(V)$ and $N(0,V,T) = N_{\infty}(V)
 \propto f_{SC}(V)$, the superconducting fraction,
 for $T > T_c$.
If we approximate $w_{dos} \propto N(0)$, then 
at a given $T < T_c^0$, the $V < V_c(T)$ samples
have $w_{dos} \sim 0$, and the $V > V_c(T)$ 
samples have $w_{dos} \sim N_{\infty}(V)$, with 
$d N_{\infty}(V)/dV < 0$. The peak would be
obviously at $V \sim V_c(T)$. The real $V$ 
dependence is of course 
smoother than what the crude argument 
above suggests.

This observation ties together {\it independent measurements}
on density of states, optical weight, and the spatial
character of the superconducting state over the complete
disorder-temperature window.

{\bf Discussion.} - In what follows we
discuss the microscopic basis of our results,
the limitations of our approach, 
and the connection of our results 
to experimental data.

i)~{\it Microscopic origin:}~Within our 
approach the thermal properties are
controlled by the mean value and thermal fluctuations of
the pairing field $\Delta_i$ and the `density' field
$\phi_i$. Disorder creates inhomogeneities 
in $n_i$ which are fed back via the Hartree shift $\phi_i$
leading to an amplified effective disorder 
$V^i_{eff}= \epsilon_i + \phi_i$.
Transport and spectral features are 
determined by the combined effect
of $V_{eff}^i$ and scattering from the $\Delta_i$.
a)~At weak disorder, $V \lesssim 0.25V_{c}$, 
the ground state has almost homogeneous $n_{i}$
and $\vert \Delta_i \vert$, with perfect
phase correlation.  With increasing $T$,
the $\langle \vert \Delta_{i} \vert \rangle$ 
increase and the $\theta_i$ randomise.
The growing amplitude and phase disorder lead
to increased scattering 
and $d \rho/dT > 0$ for $T > T_c$. 
The  $n_i$ remain roughly homogeneous
over the relevant $T$ window and the potential
scattering just adds a constant contribution to
the overall resistivity.
b)~At intermediate disorder, $0.25 V_{c} 
\lesssim V \lesssim 0.75 V_{c}$, $n_{i}$ 
is noticeably inhomogeneous in the ground
state, leading to a large $V_{eff}^i$, but
homogenize with increasing $T$.
The 
$\vert \Delta_i \vert$, on the other hand,
grow with increasing $T$ in the high temperature
regime. 
The result is a crossover in the resistivity
with increasing $T$, with the $T \gtrsim T_c$
region showing $ d \rho/dT < 0$ due
to weakening $V_{eff}^i$, and the $T \gg T_c$
region showing $d \rho/dT > 0$ due to
growing scattering from the $\Delta_i$. 
A prominent pseudogap starts to 
form below $T_{corr} \lesssim T^0_c$, where strong 
local correlations appear among $\Delta_i$, and 
deepens with decreasing $T$ until $T_{c}$, where 
the bulk superconducting transition takes place 
and a hard gap is formed.
c)~At large disorder, $V \gtrsim 0.75V_{c}$, the 
$n_i$ inhomogeneity is very large in the ground state
and SC clusters form only in `favourable' regions 
\cite{unpub} of this landscape.
The insulating regions have a larger effective gap
than the SC clusters.
The $n_i$ inhomogeneity, and the large $V_{eff}^i$ 
survives to $T \gg  T_c$ and leads to the activated 
resistivity.

ii)~{\it Limitations of our method:} There are
two sources of error in our approach.
a)~The static approximation: 
the auxiliary fields are in principle time dependent and
their temporal (quantum) fluctuations can be 
significant in the following two regimes:
1)~The vicinity of the $T=0$~~ SIT.  
As the electronic DOS becomes gapped
due to pair formation, the dynamics of these bosonic
pairs can play a significant role in transport and low
frequency optics. This has been emphasized in a very
recent preprint \cite{qxy}. However, for the $U/t=2$
that we have used, fig. 1 shows that the finite 
temperature 
SIT, for $V \lesssim 0.75V_c$, occurs in the presence
of a finite DOS at the fermi level. So, at weak to
moderate coupling, and across the finite temperature 
SIT, our approach should be useful.
2)~In the critical regime, close to $T_{c}$,
where fluctuation corrections (Aslamazov-Larkin and Maki-Thompson)
\cite{larkin-varl} are important. These also involve the dynamics of the pairing
fields, which are absent in our scheme. However, away from the immediate vicinity
of $T_c$ the effects that we highlight, 
arising from a combination
of the disorder and Hubbard interaction, would dominate.
b)~System size: weaker coupling, $U/t \lesssim 1$,
is relevant experimentally, but difficult to access with
current size limitations since the coherence length grows
as $1/\Delta_0$. Our system size, $\sim 24 \times 24$, is
significantly larger than what is accessible within QMC
but still much smaller than the inhomogeneity scales
observed experimentally. As a result, some of the 
predictions we make are only of qualitative value in
the experimental context. We discuss these next.

iii)~{\it Comparison to experiments:}~When 
comparing with experiments, it must be kept in
mind that $U=2t$ is already beyond the weak coupling 
BCS regime, with $2 \Delta / kT^0_c \sim 8$
instead of $3.5$. Most real materials, explored in the SIT 
context are however in the BCS window, so the relevant
$U_{eff}/t \lesssim 1$. This is also borne out by
the rather low $T_c^0 \sim 10$K of 
these materials \cite{sacepe-prl,pratap-ph-dg}.
 Additionally, Coulomb effects are 
neglected in our model. We touch upon 
this in the next section.
For the different indicators that we have 
calculated, a comparison to experiments reveal the 
following: 
a)~Resistivity: While our observation of 
a metal to insulator crossover in the normal state
transport is consistent with experiments, experimental
resistivities are less insulating than we observe
\cite{pratap-ph-dg,kapit-res,baturina-res}. For instance,
at $V=0.75V_{c}$, with $T_{c}=0.4T^0_c$, our resistivity 
already shows insulating behaviour, falling to one-fourth 
of its maximum value by $T=3T^0_c$, while the two 
dimensional TiN 
sample in \cite{sacepe-prl} only falls
to $60 \%$ of its maximum value even though 
$T_c \sim 0.1T^0_c$ for that sample. Similar behaviour
is seen in three dimensional NbN samples \cite{pratap-ph-dg}.
b)~Density of states:
Pseudogap effects are more pronounced in our case,
extending to larger temperatures. For instance, at $V=0.5V_{c}$,
this scale is around $T = 2.5T^0_c$, and by $V=V_{c}$, it extends
beyond $T=4T^0_c$. In contrast, 
experiments on three dimensional systems \cite{pratap-ph-dg} 
indicate that the pseudogap vanishes at a temperature
$T^*$ that decreases initially with increasing 
disorder, and finally becomes constant at $T^*  \sim 0.5T^0_c$
at large disorder. Two dimensional systems \cite{sacepe-natc}, 
on the other hand, do seem to show a qualitatively similar 
increase as ours, though their relevant scales are much smaller
(for instance at $T_{c}=0.1 T^0_c$, $T^* \sim 1.4T^0_c$, whereas
our scale is greater than $4T^0_c$.)
c)~Optics:
Low frequency features are sharply depressed,
at low $T$, in our model at large disorder, whereas there is 
a much more modest effect in experiments 
\cite{pratap-sf,crane-sf}. 
d)~Spatial character: Our results for
local indicators reveal significant similarity, but 
also points of difference, in terms of spatial
dependence and thermal variation, 
with recent experiments \cite{pratap-spat}. 
They will be dealt with
in a separate publication \cite{unpub}.

iv)~{\it Coulomb interactions:} Coulomb interactions can
affect superconductivity in two ways: 1)~weakening
of the effective pairing interaction \cite{finkel} 
and 2)~an increase in phase fluctuations \cite{em-kiv}.
The first effect can lead to an SIT driven by a vanishing gap.
However, most recent experiments suggest that the 
SIT is driven by phase fluctuations in a fragmented ground state,
and not so much by a vanishing gap. In this sense, our model captures
the correct phenomenology.
In terms of measurables, 
Coulomb effects cause a characteristic
wide dip in the DOS \cite{ar-alt}, which is absent
from our results.
They also lead to 
a contribution similar to the weak localisation
correction to the conductivity \cite{ar-alt,ar-lee}. 

{\it Conclusion:} We have studied
the transport and spectral
characteristics of a disordered s-wave superconductor
over the complete disorder
and temperature window relevant for the superconductor-insulator
transition. We have identified the 
metal to insulator crossover in the normal state
with increasing disorder and demonstrate a 
disorder window where the 
superconductor arises out of a high temperature
`insulating' state.
We map out the disorder and temperature dependence
of the single particle and optical spectra, 
discover that their low frequency weight 
is non monotonic with disorder,
and relate the weight to the 
superconducting spatial fraction in the 
disordered ground state.

{\it Acknowledgments:}
We acknowledge use of the High Performance Computing Cluster 
at HRI.  PM acknowledges support from a DAE-SRC 
Outstanding Research Investigator Award.
We thank Amit Ghosal and Pratap Raychaudhuri
 for discussions.


\begin{thebibliography}{99}
\bibitem{anderson} 
P. W. Anderson, J. Phys. Chem. Solids {\bf 11}, 26 (1959).
\bibitem{sit-revs} For reviews, see
A. M. Goldman and N. Markovic, Phys. Today {\bf 51}, No 11, 39 (1998),
V. F. Gantmakher and V. T. Dolgopolov, Phys. Usp. {\bf 53}, 3-53 (2010),
M. Sadovskii, Phys. Rep. 282, 225 (1997),
D. Belitz and T. Kirkpatrick, Rev. Mod. Phys. {\bf 66}, 261 (1994).
\bibitem{sacepe-natc} 
B. Sacepe, {\it et al.},
Nature Commun. {\bf 1}, 140 (2010).
\bibitem{sacepe-loc} 
B. Sacepe, {\it et al.},
Nature Phys. {\bf 7}, 239 (2011).
\bibitem{sacepe-prl} 
B. Sacepe, {\it et al.},
Phys. Rev. Lett. {\bf 101}, 157006 (2008).
\bibitem{pratap-ph-dg} 
M. Chand, {\it et al.},
Phys. Rev. B {\bf 85}, 014508 (2012).
\bibitem{chocka}  S.~P.~Chockalingam, {\it et al.} 
Phys. Rev. B {\bf 77}, 214503 (2008),
S.~P.~Chockalingam, {\it et al.} Phys. Rev. B {\bf 79}, 094509 (2009)
\bibitem{pratap-prl} 
M. Mondal, {\it et al.},
 Phys. Rev. Lett. {\bf 106}, 047001 (2011).
\bibitem{pratap-spat} 
A.~Kamlapure, {\it et al.}, Sci. Rep. {\bf 3}, 2979 (2013).  
\bibitem{noat} Y. Noat, {\it et al.},
Phys. Rev. B, {\bf 88}, 014503 (2013).
\bibitem{kapit-res}
 M.A.~Steiner, N.P.~Breznay, A.~Kapitulnik, Phys. Rev. B {\bf 77} 212501 (2008).
\bibitem{baturina-res}  
T.I.~Baturina, {\it et al.}, Phys. Rev. Lett. {\bf 98} 127003 (2007).
\bibitem{baturina-prl} 
T. I. Baturina, {\it et al.},
 Phys. Rev. Lett. {\bf 99}, 257003 (2007).
\bibitem{adams-mr}  V.~Yu.~Butko, P.W.~Adams, Nature {\bf 409} 161 (2001).
\bibitem{pratap-sf} M. Mondal, {\it et. al.},
 Scientific Reports {\bf 3}, 1357 (2013).
\bibitem{crane-sf} R.W.~Crane, {\it et. al.}, Phys. Rev. B, {\bf 75}, 184530 (2007).
\bibitem{ghosal} A. Ghoshal, M. Randeria and N. Trivedi, 
Phys. Rev. Lett. {\bf 81}, 3940 (1998),
A. Ghoshal, M. Randeria and N. Trivedi,
Phy. Rev. B {\bf 65}, 014501 (2001).
\bibitem{scal-dos}
C. Huscroft and R.T. Scalettar, Phys. Rev. Lett. {\bf 81}, 2775 (1998).
\bibitem{scal-res} 
R. Scalettar, N. Trivedi and C. Huscroft, Phys. Rev. B {\bf 59}, 4364 (1999).
\bibitem{bouad} K. Bouadim, {\it et. al.}, Nat. Phys. {\bf 7}, 884 (2011).
\bibitem{altland}
{\it Condensed Matter Field Theory}, A. Altland and B. D. Simons,
Cambridge University Press, (2010).
\bibitem{erez}
A. Erez and Y. Meir, Europhys. Lett. {\bf 91}, 47003 (2010).
\bibitem{dubi} Y.~Dubi, {\it et al.}, Nature, {\bf 449}, 876 (2007).
\bibitem{dag} M.~Mayr, {\it et al.}, Phys. Rev. Lett. {\bf 94}, 217001 (2005).
\bibitem{tca} S.~Kumar and P.~Majumdar, Eur. Phys. J.~B, {\bf 50}, 571 (2006).
\bibitem{rand-tc} 
T.~Paiva, {\it et. al.}, Phys. Rev. Lett. {\bf 104}, 066406 (2010).
\bibitem{unpub} S. Tarat and P. Majumdar, in preparation.
\bibitem{mbtheory} 
D. C. Mattis and J. Bardeen, Phys. Rev., {\bf 111}, 412 (1958).
\bibitem{qxy} 
M. Swanson, Y. Lee Loh, M. Randeria, N. Trivedi, 
arXiv:1310.1073.
\bibitem{larkin-varl} A.~Larkin and A.~Varlamov in 
Theory Of Fluctuations In Superconductors (Clarendon Press,
Oxford 2005).

\bibitem{finkel} A.~M.~Finkelstein, Physica B {\bf 197} (1994).
\bibitem{em-kiv} V.~J.~Emery and S.~A.~Kivelson, Phys. Rev. Lett. {\bf 74}, 3253 (1995).
\bibitem{ar-alt}  B.~L.~Altshuler and A.~G.~Aronov 
in Electron-Electron Interactions in Disordered Systems, A.
L. Efros, M. Pollak, Eds. (North-Holland, Amsterdam, 1985).
\bibitem{ar-lee} B.~L.~Altshuler, A.~G.~Aronov and P.~A.~Lee, Phys. Rev. B {\bf 44}, 1288 (1980).

\end{thebibliography}
\end{document}